# Visualization Techniques for Assessing Inhomogeneities and Stress in Ti:Sapphire crystals

VOJTĚCH MILLER[1,2,*] AND KAREL ŽÍDEK[2]

[1]Technical University of Liberec, Studentská 1402/2 461 17 Liberec 1, Liberec, Czech Republic
[2]CRYTUR, spol. s r.o., Na Lukách 2283, 511 01 Turnov, Czech Republic
[3] Institute of Plasma Physics of the Czech Academy of Sciences, Za Slovankou 1782/3, 182 00 Prague, Czech Republic
*vojtech.miller@tul.cz

**Abstract:** This study presents a comprehensive optical analysis of titanium-doped sapphire (Ti:Sa) crystals, introducing two innovative measurement techniques to enhance the characterization of this material. The first method enables highly precise transmission measurements, facilitating the visualization of optical doping patterns across samples and providing accurate Figure of Merit (FoM) evaluations. This technique covers an area of 25 × 100 mm, allowing for the creation of detailed optical property maps. The second method is specifically designed to identify stress and refractive index inhomogeneities using circular polarization, leveraging the birefringent properties of Ti:Sa material.

Experimental validation was performed on three Ti:Sa samples with distinct defects, analyzing and comparing their optical and structural properties. A novel central optical pattern, previously unreported, was observed in all samples. This pattern is hypothesized to originate from core formation during crystal growth. These findings provide new insights into the material's internal structure and hold significant implications for its optimization in optical applications.

## 1. Introduction

Titanium-doped sapphire (Ti:Sa) crystals are widely regarded as among the most versatile and efficient laser gain materials, particularly valued for their broad tunable range, spanning 650 nm to 1100 nm, and their high-power output.[1] These crystals are integral to a wide range of laser systems, where their performance is critical. However, inhomogeneities and internal stress within Ti:Sa crystals can significantly degrade laser performance, leading to reduced efficiency, beam distortion, and, ultimately, a shortened laser lifespan. [2–4] For instance, under high pump powers or high repetition rates, Ti:Sa crystals often experience uneven heating, resulting in stress fractures, optical distortions, and thermal lensing effects. [5, 6] Additionally, imperfections within the crystal can interact with high-energy pulses, further amplifying thermal and mechanical stresses. [5, 7, 8] Identifying and addressing potential defects in the bulk material before manufacturing laser components is crucial, as this can significantly reduce costs for manufacturers and improve the reliability of final products. [2, 3, 6, 9, 10]

The birefringent nature of Ti:Sa adds further complexity to the visualization and evaluation of internal defects. Conventional crossed-polarizer inspections can effectively reveal macro-defects, such as bubbles and cracks. However, due to Ti:Sa's inherent birefringence, these commonly used methods often fail to detect the distribution of internal stress, which can, for example, affect the transmitted wavefront error of manufactured elements. [11, 12] This limitation underscores the need for new methods capable of accurately assessing the quality and internal structure of Ti:Sa crystals.

In this study, we investigate the homogeneity of the optical properties and residual mechanical stress in Ti:Sa crystals. We employed an upgraded, sample-scanning version of our high-precision setup for transmission measurement. [13] This setup enabled us to study the distribution of the Figure of Merit (FoM) across crystals with varying parameters. [14]

To better understand the causes and implications of the acquired maps, we introduced a second experimental technique to visualize stress within Ti:Sa and other birefringent materials. This method employs circularly polarized light combined with polarization imaging to analyze internal stress and structural inhomogeneities. It enables the detection of subtle variations in polarization states indicative of internal stresses and defects that are otherwise challenging to observe. While this approach has been reported previously for other use cases, we adapted it to address the specific characteristics of Ti:Sa crystals. [15–18]

Additionally, we employed interferometry to map the crystals. Unlike polarization-based study, this method is sensitive to local variations in the refractive index and is unaffected by the birefringence of the crystal.

By integrating these complementary methods, we created a comprehensive framework for evaluating both the optical and structural quality of Ti:Sa crystals and interpreting 2D transmission maps from previously introduced experiments. To demonstrate this capability, we present case studies of three crystals exhibiting specific defects or inhomogeneities and describe how these defects influence the results across the applied experimental techniques.

## 2. Materials and Methods

**Materials**

The Ti:Sa crystals used in this study were sourced from Crytur spol. s. r. o. which has more than ten years of experience with Ti:Sa crystal preparation for ultrafast laser applications. Studied crystals had a diameter of approximately 55 mm, as shown in Figure 1. The lengths of the samples varied depending on the titanium doping concentration. The crystals were doped with titanium at concentrations ranging from 0.08 to 0.30 wt% of $Ti_2O_3$. All crystal samples were grown using the Czochralski method, were pulled in an a-axis direction and underwent post-growth annealing to minimize $Ti^{4+}$ concentration and reduce parasitic absorption in red-infrared region. Chosen samples were from leftovers of the produced crystals with specific defects intentionally selected for visualization and analysis in this work.

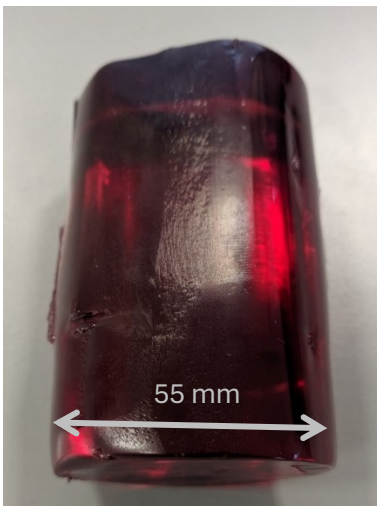

*Figure 1: Ti:Sa crystal grown by Czochralski method with cut off top and bottom parts*

**Ti:Sa FoM measurements**

The FoM calculation for Ti:Sa crystals depends on the absorption cross-section at the pump wavelength, which reflects the crystal's ability to generate inversion, and the residual absorption cross-section at the lasing wavelength, which opposes light amplification and induces crystal heating. FoM is expressed as the ratio of the absorption coefficient at the pump wavelength (commonly 532 nm) to that at the lasing wavelength (typically around 800 nm) as can be seen in eq. 1. In this work, 780 nm was used due to the availability of the laser source:

$$FoM = \frac{\alpha_{532nm} - \epsilon_{1560nm}}{\alpha_{780nm} - \epsilon_{1560nm}} \quad (1)$$

$$\epsilon_{1560nm} = \alpha_{1560nm} + \sigma_{1560nm} \quad (2)$$

Each absorption coefficient was calculated based on the sample's transmission and reflection properties. Transmission was measured directly, while reflection was calculated from the refractive index to the second order, as shown in eq. 3.

The FoM computation incorporates extinction coefficient $\epsilon_{1560nm}$ shown in eq. 2 as a reference to suppress noise introduced by non-material-related artifacts, such as bubbles or microbubbles during the scanning process which are included in the scattering losses $\sigma_{1560nm}$. While Ti:Sa crystals exhibit distinctive absorption peaks in the visible range (around 490 nm) and near-infrared range (around 800 nm), no significant peaks are observed further into the NIR range, making 1560 nm an appropriate reference wavelength. The spectral characteristics of Ti:Sa have been thoroughly studied by Gong et al. [19]

**FoM visualization**

Figure 2 illustrates the experimental setup used for measuring the Figure of Merit (FoM), as previously reported in [13]. In contrast to the earlier experiment, the setup was upgraded to enable 2D measurements, allowing analysis in both directions perpendicular to the crystal axis.

The experimental setup utilized three distinct wavelengths produced by two lasers – the second harmonic generation (SHG) of Nd:YAG laser (532 nm) combined with its fundamental beam and SHG of Er:YAG laser (780 and 1560 nm). The laser beams originated from two input branches. The first branch consisted of Nd:YAG laser, where the SHG was utilized, and the IR filter was applied to suppress residual 1064 nm light. The second input branch employed of Er:YAG laser (Toptica Photonics FemtoFErb 1560), which was used at its fundamental wavelength (1560 nm) and combined with its second harmonic (780 nm) generated by an SHG crystal (PPLN, Covesion).

All three beams were aligned to propagate along a common beamline through the remainder of the setup. First, their polarization was precisely adjusted to purely horizontal using a Rochon prism, achieving a polarization degree of $1:10^5$. The horizontally polarized beams were then split into two branches using a beam splitter. The first branch (denoted as the sample branch) passed through the sample, while the second branch (denoted as the reference branch) was used to compensate for fluctuations in laser intensity. Both branches passed through a dual-channel optical chopper blade (ratio 3:7) to modulate their intensity.

The intensities of the sample and reference beams were measured using a pair of integrating spheres, each equipped with photodetectors dedicated to the three wavelengths. Specifically, two Thorlabs DET36A2 photodiodes were used for 532 nm and 780 nm, and one Newport Large Area Photodetector Model 2317 was employed for 1560 nm. The signals from the photodiodes were amplified and digitally processed using Fourier transformation to extract the intensity of both beams. A key advantage of this setup was the ability to use the same detector to measure both the sample and reference beam intensities, as they were distinguishable by their modulation via the optical chopper.

The sample stage was mounted on a set of Thorlabs motorized 1D stages, enabling sample movement for 2D scanning across a maximum area of 100 × 25 mm. Given the crystal diameter of approximately 55 mm, the typical scanned area was about 60 × 25 mm. This allowed the creation of detailed visualization maps that could be directly compared with other visualization techniques.

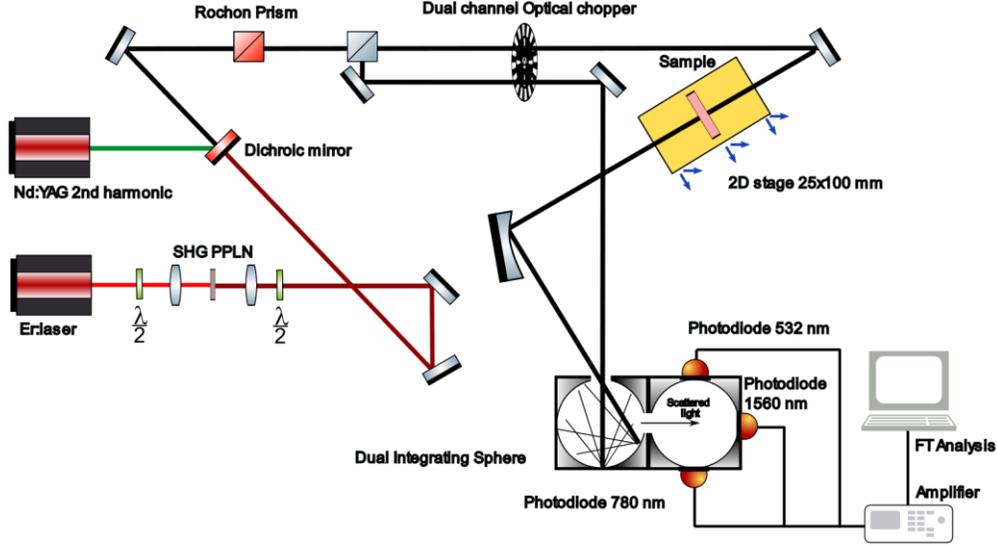

*Figure 2: Schematics of FoM visualization experiment*

During the measurement process, each laser (532 nm, 780 nm, and 1560 nm) was sequentially used to illuminate the Ti:Sa crystal. Transmission measurements were recorded using the photodetectors as described in the previous section. The recorded transmission values, combined with the measured thickness of the samples, were used to calculate the absorption maps using eq. 3

$$\alpha_{i[j,k]} = \frac{ln\left(\frac{(1 - R_{p_i})^2 + R_{p_{i[j,k]}}^2}{T_{i[j,k]}}\right)}{d} \quad (3)$$

Where $\alpha_{i[j,k]}$ is absorption coefficient at given wavelength i at pixel position j,k. Rp corresponds to a reflection coefficient for p-polarized light. $T_{i[j,k]}$ is transmission at given wavelength i at pixel position j,k and d is sample thickness.

Absorption maps were derived from the transmission maps by accounting for the first two orders of reflection for each wavelength. The Figure of Merit (FoM) was then calculated by dividing the absorption map at 532 nm by the absorption map at 780 nm and subsequently subtracting the absorption map at 1560 nm.

The inclusion of the absorption map at 1560 nm as a reference significantly suppressed major defects, such as cracks or bubbles, within the crystal samples. This approach allowed for a more refined assessment of fine variations in absorption and FoM across the crystal samples.

For all samples, transmission maps were measured for light polarized parallel to the c-axis of the Ti:Sa crystal. Diameter of used beams ranged from 1 to 1.5 mm based on wavelength used.

**Stress visualization**

Figure 3 shows the experimental setup for analyzing refractive index inhomogeneities and internal stress in birefringent samples. Circularly polarized light was generated by passing a linearly polarized beam from a 663 nm red laser through a quarter-wave (λ/4) plate. The beam was expanded to cover approximately 80% of the Ti:Sa crystal samples, which typically had a diameter of 55 mm. Using 2-inch optics, the beam was expanded to a diameter of roughly 45 mm and then de-expanded after passing through the sample to match the detector size of the polarization camera (14 × 10 mm).

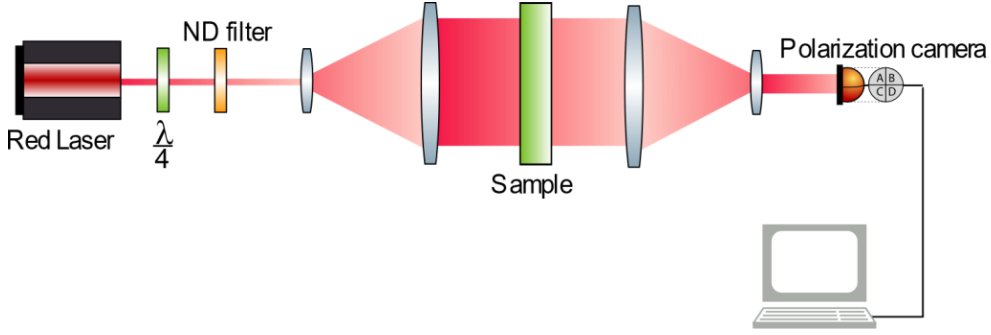

*Figure 3: Experiment to visualize index of refraction inhomogeneities (internal stress)*

The polarization camera captured light intensity in four polarization states simultaneously, using a linear polarization filter in front of each super-pixel detector (0°, 45°, 90°, and 135°). These intensities were used to calculate the phase difference $\Delta$ and azimuthal angle $\phi$ within the sample using eq. 4 and eq. 5:

$$\Delta = sin^{-1}\left(\frac{\sqrt{(I_3 - I_1)^2 + (I_2 - I_4)^2}}{I_0}\right) \quad (4)$$

$$\varphi = \frac{1}{2} tan^{-1}\left(\frac{I_3 - I_1}{I_2 - I_4}\right) \quad (5)$$

The resulting phase difference ranges from 0 to π/2 radians, while the azimuthal angle spans ±90° [18].

Images of the phase differences within the samples were examined and compared with other measuring techniques. The primary goal was to visualize the core of the crystal and possible remaining mechanical stress present from either growing or processing of the crystal.

To establish a baseline for stress visualization, a quartz sample was used as a reference. Measurements were taken both without induced pressure and with applied pressure by tightening a flat screw from the side of the sample. These reference measurements, shown in Figure 4, validate the method by confirming its ability to detect stress accurately.

From the image analysis, a phase difference Δ of approximately 0.3 rad was observed. Using eq. 6, the stress applied on the sample was calculated as follows:

$$\sigma_1 = \sigma_2 + \frac{\Delta\lambda}{2\pi t C} \quad (6)$$

Here $\sigma_1$ and $\sigma_2$ represent the stress along the principal axes. It was assumed that $\sigma_2 = 0$. The wavelength $\lambda$ of the light source was 663 nm, thickness *t* of the sample was 2 mm and stress-optic coefficient *C* for fused silica was 2.43 ·10$^{-12}$ Pa$^{-1}$. [20]

The resulting induced stress, $\sigma_1$, was calculated as 22 MPa, consistent with expectations for the applied screw pressure.

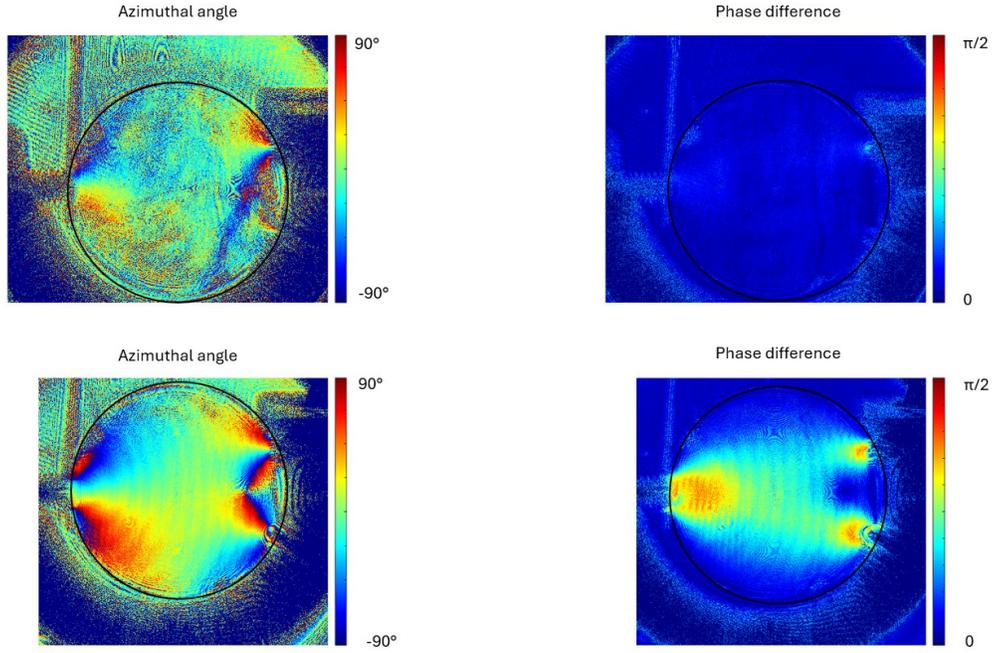

*Figure 4: Birefringence of a quartz substrate. Upper row: data without load; bottom row: data with load added by tightening a flat screw from the left side. Left panels: azimuthal angle, right panel: phase difference. The sample position is marked by black circles in each image, the diameter of the sample is 1 inch. The substrate was placed in a holder ("c" shaped dark feature below).*

To validate the methods, three distinct Ti:Sa samples with varying defects were measured using the newly introduced experimental setups as well as an interferometer. This allowed for comparison, correlation of results, and interpretation of the observed features.

In the following subsections, we will first describe the three testing crystals and present the measured maps, followed by detailed discussions of the results, overviewed in Table 1 and Figure 5.

**Table 1: Overview of the measured Ti:Sa samples.**

| Sample | Thickness [mm] | Abs. coef. @532 nm [cm$^{-1}$] | FoM | Defect(s) |
|---|---|---|---|---|
| #1 | 13.3 | 5.3 | 200 | Sub-surface scatter |
| #2 | 15.7 | 2.3 | 150 | Microbubbles, 2x deep scratch |
| #3 | 27.1 | 1.85 | 500 | Microbubbles in the center |

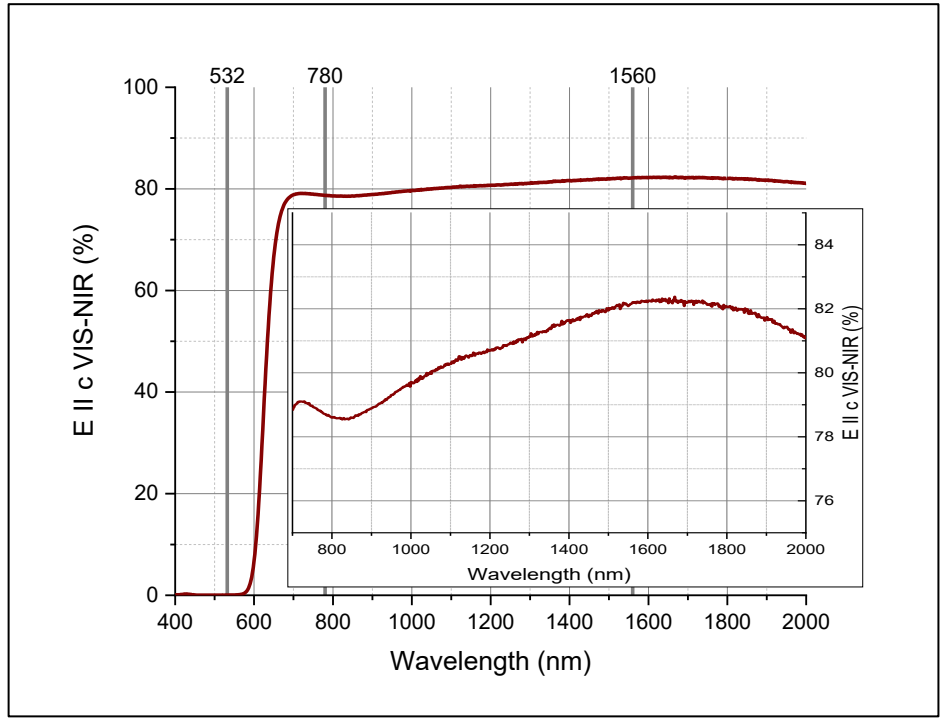

*Figure 5: Spectrum of Ti:Sa material (Sample #1) for E ∥ c polarization. Region 700-2000 nm plotted to secondary axis for better visualization. Grey vertical lines represent 532, 780 and 1560 nm regions. Typical residual absorption caused by $Ti^{4+}$ ion can be seen around 800 nm. Near-zero transmission in VIS region is caused by thickness of the sample.*

**Interferometric measurement**

Interferometric measurements were performed using a low-coherence interferometer (Optoflat, Interoptics). The measurements focused exclusively on visualizing internal defects within the samples. This was achieved by capturing a series of interferograms and subsequently subtracting them to isolate material-related features while eliminating surface deformation effects.

This measurement regime proved advantageous as polished samples exhibited certain wavefront deformations on both surfaces that could have adversely affected the interferometric maps. By removing these surface artifacts, the resulting maps provided a clearer representation of internal defects.

## 3. Results

**Sample #1**

*Observation*

Sample #1 exhibited a prominent subsurface scattering plane located less than 1 mm beneath the surface likely consisting of periodical micro-inclusions. This defect scattered incident light and was visible under white light illumination (*Figure 6*). The defect spanned approximately half the sample's area. While such a defect is easily avoidable in actual processing it posed challenges for evaluating spectroscopic properties like the FoM and absorption coefficients.

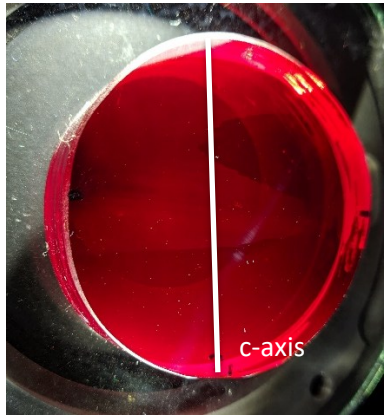

*Figure 6: Ti:Sa sample #1 - side lit by white light, approximate diameter of 55 mm. The added line provides a reference orientation for Figures 6-10.*

### Interferometric Results

The Total Wavefront Error (TWE) measurement (*Figure 7*) highlighted the boundary between the defect and the rest of the sample. Other observed defects were on a significantly smaller scale.

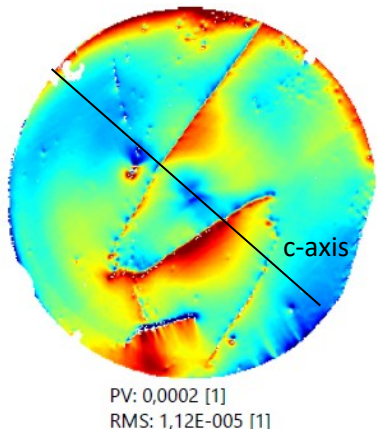

*Figure 7: Ti:Sa sample #1 – interferometer image, used wavelength 632 nm, automatic detection of the sample boundaries – 95 % of the sample area. The added line provides a reference orientation for Figures 6-10. PV values represent difference in index of refraction throughout the sample which has dimensionless units.*

### Transmission and Absorption Mapping

*Figure 8* presents transmission maps at 532 nm, 780 nm, and 1560 nm wavelengths, along with the corresponding absorption coefficients. The defect region was prominent at 780 nm and 1560 nm but obscured at 532 nm. The absorption coefficient at 532 nm was approximately 5.3 $cm^{-1}$, aligning with expectations for this sample, which had the highest titanium concentration among those tested.

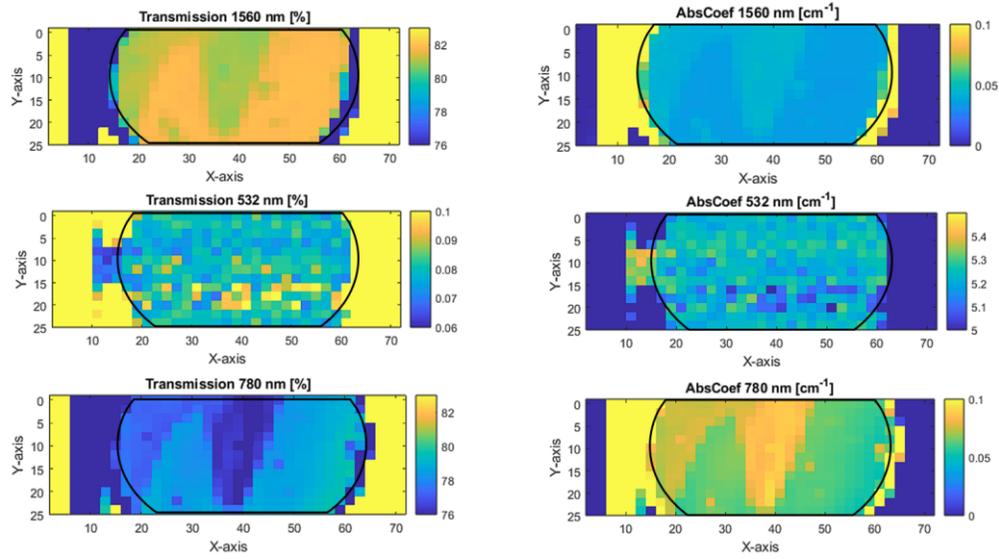

*Figure 8: Ti:Sa sample #1 – Left panels: transmission maps on the wavelength 532 nm, 780 nm and 1560 nm. Right panels: corresponding absorption coefficients on the respective wavelengths. Black lines show the boundary of the sample. C-axis oriented horizontally.*

### FoM and Phase Mapping

The FoM map (*Figure 9*) derived from the absorption maps, emphasizes the impact of the subsurface defect.

On a phase map (*Figure 10*), a notable phase change was observed along the defect's border. While slight central variations were detected, other areas of the sample showed no significant defects. Interference fringes from the coherent light source partly obscured the phase map, a recurring issue across all samples.

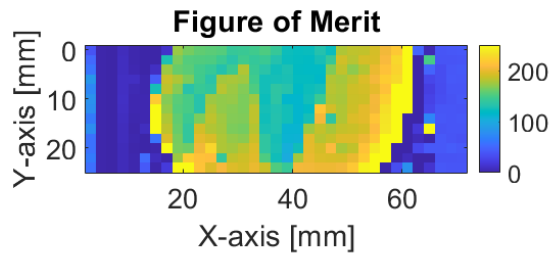

*Figure 9: Ti:Sa sample #1 – FoM map. C-axis oriented horizontally.*

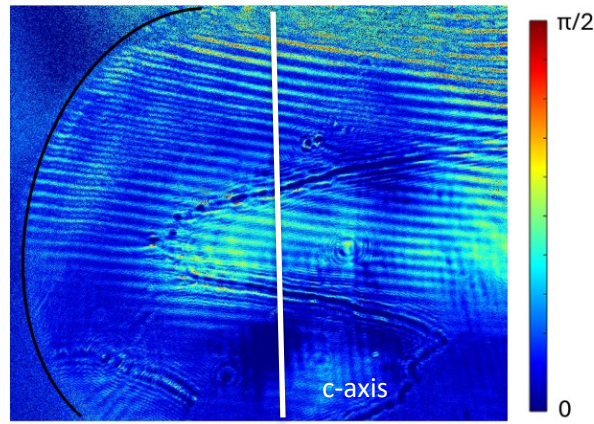

*Figure 10: Ti:Sa sample #1 – phase difference map, black line depicts edge of the sample*

**Sample #2**

*Observation*

Sample #2 featured two noticeable surface scratches, one of which was accompanied by a minor microbubble region. Under white light illumination (*Figure 11*), the sample appeared mostly clear apart from these features. The absorption coefficient at 532 nm was approximately 2.3 cm$^{-1}$.

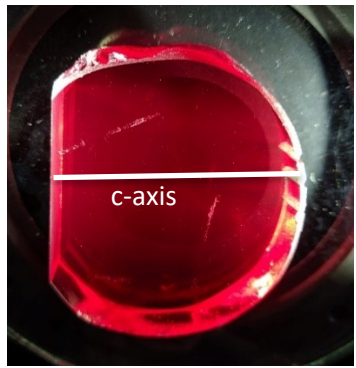

*Figure 11: Ti:Sa sample #2 – side lit by white light, approximate diameter of 55 mm. The added line provides a reference orientation for Figures 11-15.*

*Interferometric Results*

The interferometric map (*Figure 12*) revealed an asymmetric feature in the central area, corresponding to refractive index inhomogeneities.

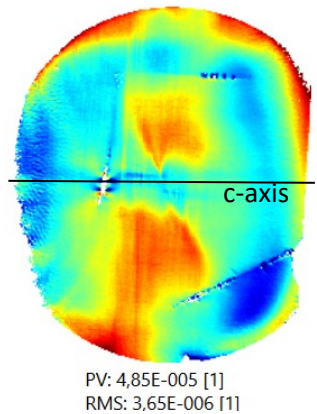

*Figure 12: Ti:Sa sample #2 – interferometer image, used wavelength 632 nm, automatic detection of the sample boundaries – 95 % of the sample area. PV values represent difference in index of refraction throughout the sample which has dimensionless units.*

### Transmission and Absorption Mapping

*Figure 13* illustrates the transmission and absorption maps, showing a central nonsymmetrical distribution at 532 nm, which we will hereafter refer to as "hourglass"-shaped distribution. This feature, also visible in the absorption map, will be further analyzed in the discussion section.

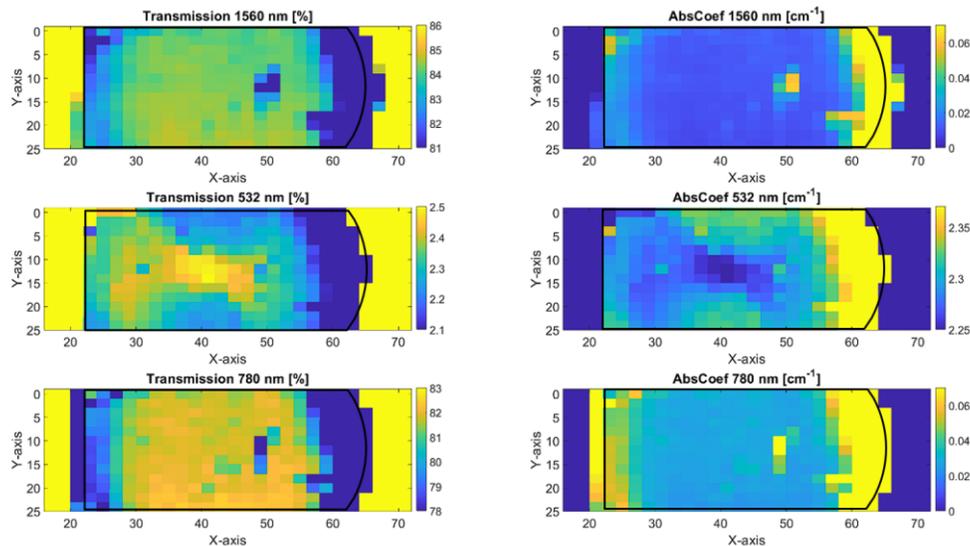

*Figure 13: Ti:Sa sample #2 – Left panels: transmission maps on the wavelength 532 nm, 780 nm and 1560 nm. Right panels: corresponding absorption coefficients on the respective wavelengths. Black lines show the boundary of the sample. C-axis oriented horizontally.*

### FoM and Phase Mapping

The resulting FoM Map (*Figure 14*) was only mildly influenced by the central bubble region.

The Phase Map (*Figure 15*) revealed both scratches and a distinct phase-shifted region on the right side of the sample.

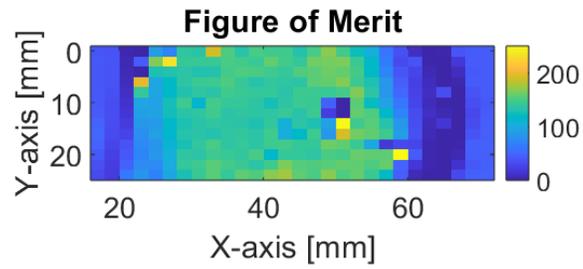

*Figure 14: Ti:Sa sample #2 - FoM map. C-axis oriented horizontally.*

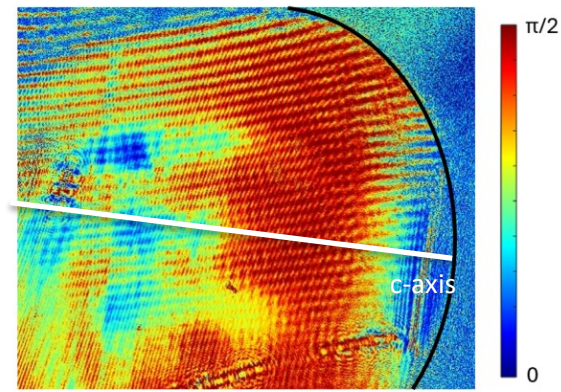

*Figure 15: Ti:Sa sample #2 – phase difference map, black line depicts edge of the sample*

**Sample #3**

*Observation*

Sample #3 had a single subtle microbubble region in the center, with no other defects detected under white light illumination (*Figure 16*). The absorption coefficient at 532 nm was approximately 1.85 cm$^{-1}$.

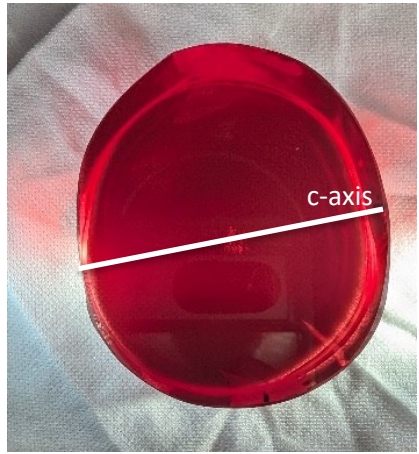

*Figure 16: Ti:Sa sample #3 – side lit by white light, approximate diameter of 55 mm. The added line provides a reference orientation for Figures 16-20.*

### Interferometric Results

The interferometric map (*Figure 17*) revealed "hourglass"-shaped wavefront defects in the central region, analogous to those observed in Sample #2.

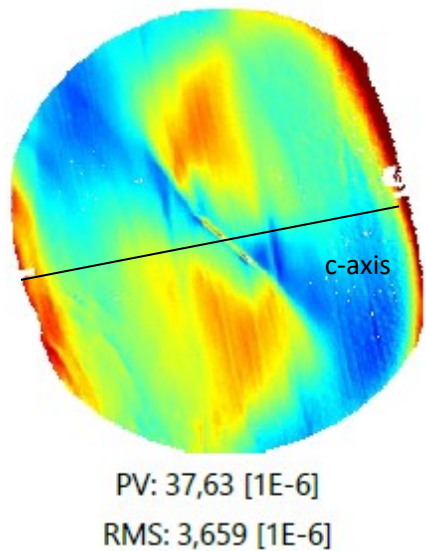

PV: 37,63 [1E-6]
RMS: 3,659 [1E-6]

*Figure 17: Ti:Sa sample #3 – interferometer image, used wavelength 632 nm, automatic detection of the sample boundaries – 95 % of the sample area. PV values represent difference in index of refraction throughout the sample which has dimensionless units.*

### Transmission and Absorption Mapping

Interestingly, the absorption maps in *Figure 18* show similar structure as with Sample #2. The region of microbubbles in the central part is again visible at 780 nm and 1560 nm. On 532 nm there is a clear centrally symmetrical distribution of absorption across the sample. This aligns with the structure that can be seen in the interferogram above and phase map below.

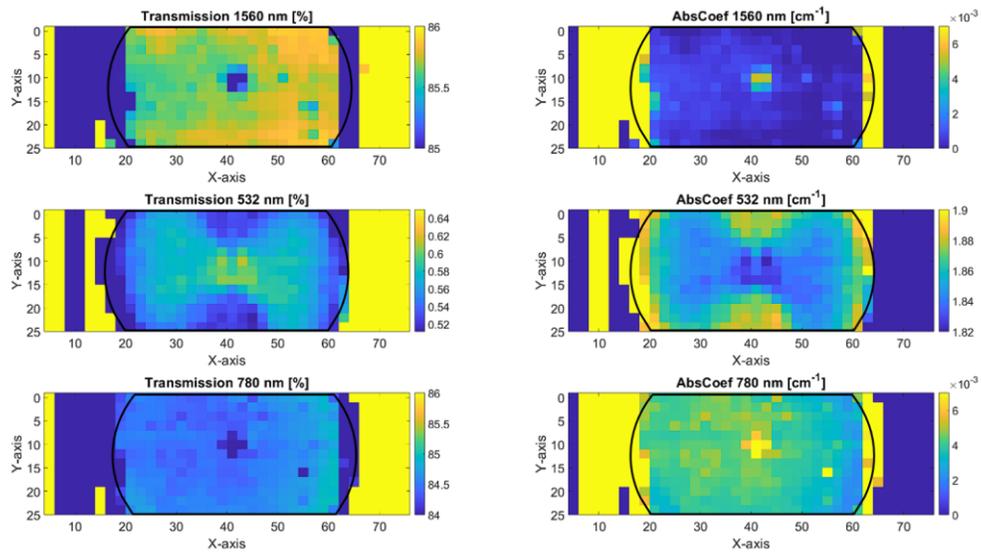

*Figure 18: Ti:Sa sample #3 – Left panels: transmission maps on the wavelength 532 nm, 780 nm and 1560 nm. Right panels: corresponding absorption coefficients on the respective wavelengths. Black lines show the boundary of the sample. C-axis oriented horizontally.*

### FoM and Phase Mapping

The FoM Map (*Figure 19*) is dominated by inhomogeneities in the IR region, reflecting the observed structural features.

The Phase Map (*Figure 20*) which shows tilted sample revealed a distinct "hourglass"-shaped phase change in the center, corresponding to the 532 nm absorption asymmetry. A local phase change along the left edge suggested the presence of a microcrack, with stress fading toward the center.

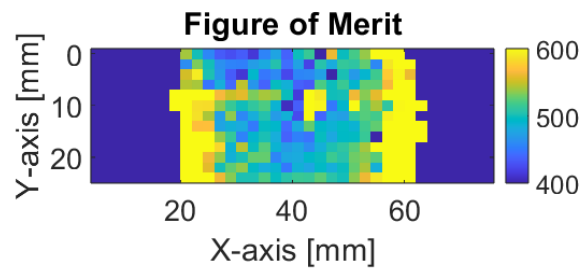

*Figure 19: Ti:Sa sample #3 – FoM map. C-axis oriented horizontally.*

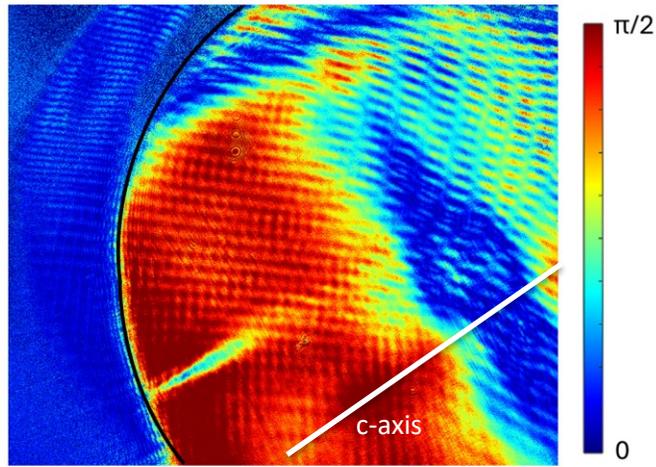

*Figure 20: Ti:Sa sample #3 – phase difference map, black line depicts edge of the sample*

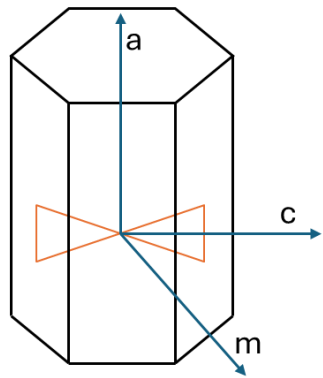

Figure 21: Hourglass shape relative to the crystal lattice of Ti:Sa

### 4. Discussion

The presented measurements were conducted on three Ti:Sa samples, each containing specific defects. These defects were first identified through preliminary examination using both the naked eye and white light, which revealed general surface imperfections and inhomogeneities. More detailed insights into the internal structure of each sample were provided by the subsequent methods employed in this study. Each of these complementary techniques helped create a comprehensive understanding of the defects and stress distribution in the Ti:Sa crystals.

The transmission measurement setup was designed with high precision to detect the variations in high-absorbing and low-absorbing regions, where a variation in the absorption coefficient corresponds to a slight change in the transmission coefficient. This is essential for the Ti:Sa samples with a high FoM value, where the measured wavelengths correspond to this case. The observed range of optical density for 532 nm in the samples has been 0.1 cm$^{-1}$ with distinguishable features in the variations of 0.01 cm$^{-1}$.

Owing to the precision, we were able to observe "hourglass"-shaped feature in absorption at 532 nm wavelength in Samples #2 and #3, a phenomenon that has not been documented in previous literature. The regions of varying optical density reflect the local concentration of the Ti$^{3+}$ ions.

Optical density at the lasing wavelength (780 nm) reflects the concentration of $Ti^{4+}$ ions. The maps lack the central feature, and the local absorption coefficient was instead dominated by the inhomogeneous scattering losses. This could be confirmed by observing the same high-scattering regions at 1560 nm, where the Ti:Sa is expected to be transparent. [4, 17]

However, the central feature is apparent in both Sample #2 and #3 in the interferometric measurement, which indicates the local variation in the refractive index. Low phase shift, i.e. lower refractive index, corresponds to the axis with higher doping – the m-axis, visualized in Figure 21 . We interpret this asymmetry to arise from the core formation of the Ti:Sa crystal as all other factors involved in the crystal growth process are axially symmetrical. The Ti:Sa crystal lattice orientation is defined by the seeding crystal, which was oriented along the a-axis, perpendicular to the a-plane {1120} [21].  The crystal growth follows the direction of the a-axis. The core formation should result in localized variations in the dopant concentration and, subsequently, stress, leading to the observed asymmetry. This interpretation is based on the parallel of doped YAG crystals where the core also introduces doping asymmetry and stress pattern in both radial and axial planes. [22]

The visualization of stress in Sample 1 clearly shows the defect boundary. When we compare the PV of Sample #1 to other measured samples, we can see that PV value is almost one order of magnitude higher than in the remaining samples.  These additional dominant losses in Sample #1 prevent us from determining, whether the "hourglass"-shaped asymmetry is also present here. The transmittance reveals regions with the increased scattering, which are, however, not apparent in other experiments.

The most important parameter related to the crystal applications is the resulting FoM. We studied three samples with absorption coefficient at 532 nm at polarization parallel to the c-axis ranging from 1,85 to 5,3 cm$^{-1}$ at5 532 nm. The resulting FoM values were in range from 100 to 500. An overview of the data is provided in Table 1.

The FoM value of 200 outside of the defect region in Sample #1 is generally considered as a great value for such high doping of the titanium and it clearly shows a degradation of the FoM value in the defect region, primarily due to increased absorption in the 780 nm region. This suggests that the actual FoM values of the material might be similar to those in the rest of the sample once the subsurface defect is removed (either by cutting or grinding). The sample #2 shows FoM values of 150 as well but with a lower doping. When we compare the mean absorption levels on 780 nm on sample #2 and #3 we can see that the remaining absorption is approximately six times higher (0,03 cm$^{-1}$) in sample #2 than on sample #3 (0,005 cm$^{-1}$) which is the main cause for the FoM levels differences. We note that sample #3 showed superior FoM values (around 500) compared to the other samples. Despite having significantly different FoM values, both Samples 2 and 3 showed the same central asymmetric "W" feature.

## 5. Conclusion

In this study, we employed a combination of precise transmission measurements, interferometric analysis, and stress visualization techniques to investigate three Ti:Sa samples, each containing specific defects. Through these complementary methods, we gained detailed insights into the internal structure, stress distribution, and absorption characteristics of the samples. Notably, the transmission measurements revealed non-axial symmetry and a unique "hourglass" shape in the absorption distribution of samples #2 and #3. This anomaly is likely linked to core formation in the crystal lattice during the growth process, affecting the dopant distribution.

The study also highlighted significant variations in FoM values across the samples, with sample #3 showing the highest FoM, indicating superior optical performance. Meanwhile, stress visualization correlated well with dopant asymmetry observed in the absorption maps whose origin is believed to stem from core formation in the crystal. These findings underscore the need for careful monitoring of the crystal growth process and dopant distribution to optimize the performance of Ti:Sa laser gain materials. Moreover, the methods developed here offer a robust framework for future quality control and detection in Ti:Sa crystals.

## 6. Acknowledgements

**Funding.** Grantová Agentura České Republiky (GA22-09296S); Ministerstvo Školství, Mládeže a Tělovýchovy (CZ.02.1.01/0.0/0.0/16_026/0008390); Technická Univerzita v Liberci (SGS-2024-3484).

**Disclosures.** The authors declare no conflicts of interest.

**Data availability.** Data underlying the results presented in this paper are not publicly available at this time but may be obtained from the authors upon reasonable request.